\documentclass[preprint,showpacs,preprintnumbers,amsmath,amssymb]{revtex4}

\usepackage{graphicx}% Include figure files
\usepackage{dcolumn}% Align table columns on decimal point
\usepackage{bm}% bold math
\usepackage{latexsym}
\usepackage{amsmath}
\usepackage{amssymb}
\usepackage{psfrag}
\usepackage[latin1]{inputenc}
\usepackage[T1]{fontenc}

\newcommand\RR{\mathbb{R}}
\newcommand\NN{\mathbb{N}}
\newcommand\dps{\displaystyle }

\begin{document}

\preprint{A6.06.173}

\title{An efficient sampling algorithm for Variational Monte Carlo}
\author{Anthony Scemama, Tony Leli\`evre, Gabriel Stoltz, Eric Canc\`es}
\affiliation{
CERMICS and INRIA Project Micmac,
Ecole Nationale des Ponts et Chauss\'ees,   
6 et 8 avenue Blaise Pascal, Cit\'e Descartes - Champs sur Marne
77455 Marne la Vall\'ee Cedex 2, France.
}
\author{Michel Caffarel}
\affiliation{
Laboratoire de Chimie et Physique Quantiques, CNRS-UMR 5626,
IRSAMC Universit\'e Paul Sabatier, 118 route de Narbonne
31062 Toulouse Cedex, France.
}
\date{\today}

%\maketitle
%\newpage

\begin{abstract}
%\section*{Abstract}
We propose a new algorithm for sampling the $N$-body
density $|\Psi({\bf R})|^2/\int_{\RR^{3N}} |\Psi|^2$ in the
Variational Monte Carlo (VMC) framework. This algorithm is based upon a
modified Ricci-Ciccotti discretization of the Langevin dynamics in the phase
space $({\bf R},{\bf P})$
improved by a Metropolis acceptation/rejection step. We show through
some representative numerical examples (Lithium, Fluorine and Copper
atoms, and phenol molecule), that this algorithm is superior to the 
standard sampling algorithm based on the biased
random walk (importance sampling).

\end{abstract}
%\newpage

\maketitle

\section{Introduction}

Most quantities of interest in quantum physics and chemistry are
expectation values of the form
\begin{equation}
 \frac{\langle \Psi, \hat O \Psi \rangle}{\langle \Psi, \Psi
 \rangle} 
\end{equation}
where $\hat O$ is the self-adjoint operator (the observable) associated
with a physical quantity $O$ and $\Psi$ a given wave function. 

For $N$-body systems in the position representation, $\Psi$ is a
function of $3N$ real variables and
\begin{equation} \label{eq:integral}
\frac{\langle \Psi, \hat O \Psi \rangle}{\langle \Psi, \Psi
 \rangle}  = \frac{\dps 
\int_{\RR^{3N}} [\hat O\Psi]({\bf R}) \,
\Psi({\bf R})^\ast \, \text{d}{\bf R}}{\dps 
\int_{\RR^{3N}} |\Psi({\bf R})|^2 \, \text{d}{\bf R}}.
\end{equation}
High-dimensional integrals are very difficult to evaluate numerically by
standard integration rules. For specific operators $\hat O$ and specific
wave functions $\Psi$, {\it e.g.} for electronic Hamiltonians and Slater
determinants built from Gaussian atomic orbitals, the above integrals can
be calculated analytically. In some other special cases, (\ref{eq:integral}) can
be rewritten in terms of integrals on lower-dimensional spaces
(typically $\RR^3$ or $\RR^6$). 

In the general case however, the only possible way to evaluate
(\ref{eq:integral}) is to resort to stochastic techniques. The VMC
method~\cite{bressanini1} consists in remarking that
\begin{equation} \label{eq:reformulation}
\frac{\langle \Psi, \hat O \Psi \rangle}{\langle \Psi, \Psi
 \rangle}  = \frac{\dps \int_{\RR^{3N}} O_L({\bf R}) \, 
|\Psi({\bf R})|^2 \, \text{d}{\bf R}}{ \dps \int_{\RR^{3N}}  
|\Psi({\bf R})|^2 \, \text{d}{\bf R}
}
\end{equation}
with $O_L({\bf R}) = [\hat O\Psi]({\bf R})/\Psi({\bf R})$, hence
that 
\begin{equation}
\frac{\langle \Psi, \hat O \Psi \rangle}{\langle \Psi, \Psi
 \rangle} \simeq \frac{1}{L} \sum_{n=1}^L
O_L({\bf R}^n)   
\end{equation}
where $({\bf R}^n)_{n \geq 1}$ are points of $\RR^{3N}$ drawn from
the probability distribution $|\Psi({\bf R})|^2/\int_{\RR^{3N}} |\Psi|^2$. 

The VMC algorithms described in the present article are generic, in the
sense that they can be used to compute the expectation value of any
observable, for any $N$-body system. In the numerical example, we will
however focus on the important case of the calculation of electronic energies
of molecular systems. In this particular case, the expectation value to
be computed reads 
\begin{equation} \label{eq:energy}
\frac{\langle \Psi, \hat H \Psi \rangle}{\langle \Psi, \Psi
 \rangle}  = \frac{\dps \int_{\RR^{3N}} E_L({\bf R}) \, 
|\Psi({\bf R})|^2 \, \text{d}{\bf R}}{ \dps \int_{\RR^{3N}}  
|\Psi({\bf R})|^2 \, \text{d}{\bf R}
}
\end{equation}
where the scalar field $E_L({\bf R}) = [\hat
H\Psi]({\bf R})/\Psi({\bf R})$ is called the {\em local
  energy}. Remark that if $\Psi$ is an eigenfunction of $\hat H$
associated with the eigenvalue $E$, $E_L({\bf R}) = E$ for all ${\bf R}$.
Most often, VMC calculations are performed with
trial wave functions $\Psi$ that are good approximations of some ground 
state wave function $\Psi_0$. Consequently, $E_L({\bf R})$ usually is
a function of low variance (with respect to the probability density 
$|\Psi({\bf R})|^2/\int_{\RR^{3N}} |\Psi|^2$). This is the reason
why, in practice, the approximation formula
\begin{equation}
\frac{\langle \Psi, \hat H \Psi \rangle}{\langle \Psi, \Psi
 \rangle} \simeq \frac{1}{L} \sum_{n=1}^L
E_L({\bf R}^n)   
\end{equation}
is fairly accurate, even for relatively small values of $L$ (in practical
applications on realistic molecular systems $L$ ranges typically
between $10^6$ and $10^9$).

Of course, the quality of the above approximation formula depends on the
way the points $({\bf R}^n)_{n \geq 1}$ are generated. In
section~\ref{sec:RW_CS}, we describe the standard sampling method
currently used for VMC calculations. It
consists in a biased (or importance sampled) random walk in the configuration
space (also called position space) $\RR^{3N}$ corrected by a Metropolis
acceptation/rejection procedure. In section~\ref{sec:RW_PS}, we
introduce a new sampling scheme in which the points $({\bf R}^n)_{n \geq
  1}$ are the projections on the configuration space of one realization of
some Markov chain on the phase space (also called position-momentum space)
$\RR^{3N} \times \RR^{3N}$. This Markov chain is obtained by a modified
Langevin dynamics, corrected by a Metropolis acceptation/rejection
procedure. 

Finally, some numerical results are presented in
section~\ref{sec:num}. Various sampling algorithms are compared and it
is demonstrated on a bench of representative examples that the algorithm
proposed here based on the modified Langevin dynamics is the most efficient
one (the mathematical criteria for measuring the efficiency will be made
precise below).  

Before turning to the technical details, let us briefly comment on the
underlying motivations of our approach.
The reason why we have introduced a (purely fictitious) Langevin dynamics in
the VMC framework is twofold:
\begin{itemize}
\item first, sampling methods based on Langevin dynamics turn out to
  outperform those based on biased random walks in classical molecular
  dynamics (see~\cite{CLS} for a quantitative study on carbon chains); 
\item second, a specific problem encountered in VMC calculations on
  fermionic systems is that the standard discretization of the biased
  random walk (Euler scheme) does not behave properly close to the nodal
  surface of the
  trial wave function $\Psi$. This is due to the fact that the drift term
  blows up as the inverse of the distance to the
  nodal surface: if a random walker gets close to the nodal surface,
  the drift term repulses it far apart in a single time
  step. As demonstrated in~\cite{caf,UNR}, it is possible to 
  partially circumvent this difficulty by resorting to more clever
  discretization schemes. Another strategy consists in replacing the
  biased random walk by a Langevin dynamics: the walkers then have a
  mass (hence some inertia) and the singular drift does not directly act
  on the position variables (as it is the case for the biased random
  walk), but indirectly {\it via} the momentum variables. The
  undesirable effects of the singularities are thus expected to be
  damped down. 
\end{itemize}

\section{Description of the algorithms} \label{sec:algo}

\subsection{Metropolis algorithm}

The Metropolis algorithm~\cite{metropolis} is a general purpose sampling
method, which combines the simulation of a Markov chain with an
acceptation/rejection procedure. 

\medskip

In the present article, the underlying state space is  
either the configuration space $\RR^{3N}$ or the phase space
$\RR^{3N} \times \RR^{3N}  \equiv \RR^{6N}$. Recall that a Markov chain
on $\RR^d$ is characterized by 
its transition kernel $p$. It is by definition the non-negative 
function of $\RR^d \times {\mathcal B}(\RR^d)$ (${\mathcal B}(\RR^d)$ is
the set of all the Borel sets of $\RR^{d}$) such that, if ${\bf X} \in
\RR^{d}$ and $B \in {\mathcal B}(\RR^d)$, the probability for the Markov
chain to lay in $B$ at step $n+1$ if it is at ${\bf X}$ at step $n$ is
$p({\bf X},B)$. The transition kernel has a density with respect to the
Lebesgue measure if for any ${\bf X} \in \RR^d$, there exists a
non-negative 
function $f_{\bf X} \in L^1(\RR^{d})$ such that
\begin{equation}
p({\bf X},B) = \int_{B} f_{\bf X}({\bf X}') \, \text{d}{\bf X}'.
\end{equation}
The non-negative number $f_{\bf X}({\bf X}')$ is often denoted by $T({\bf X} \rightarrow
{\bf X}')$ and the function $T \; : \; \RR^d \times \RR^d \longrightarrow
\RR_+$ is called the transition density.  

\medskip

\noindent
Given a Markov chain on $\RR^d$ with transition density $T$ and a
  positive function $f \in L^1(\RR^d)$, the Metropolis
  algorithm consists in generating a sequence $({\bf X}^n)_{n \in 
  \NN}$ of points in $\RR^{d}$ starting from some point ${\bf X}^0 \in
  \RR^{d}$ according to the following iterative procedure:
  \begin{itemize}
  \item propose a move from ${\bf X}^n$ to $\widetilde {\bf X}^{n+1}$ according to
  the transition density $T({\bf X}^n \rightarrow \widetilde {\bf X}^{n+1})$;
\item compute the acceptance rate 
\begin{equation*}
 A({\bf X}^n  \rightarrow \widetilde {\bf X}^{n+1}) = \min \left(
 \frac{f(\widetilde {\bf X}^{n+1})
 \,T(\widetilde{\bf X}^{n+1} \rightarrow 
 {\bf X^{n})}} {f({\bf X}^{n}) \,T({\bf X}^{n} \rightarrow
 \widetilde{\bf X}^{n+1})},\,1
 \right);
\end{equation*}

\item draw a random variable $U^n$ uniformly distributed in $[0,1]$;
  \begin{itemize}
  \item if $U^n \le  A({\bf X}^n  \rightarrow \widetilde
  {\bf X}^{n+1})$, accept the move: ${\bf X}^{n+1}
  = \widetilde {\bf X}^{n+1}$; 
\item if $U^n >  A({\bf X}^n  \rightarrow \widetilde
  {\bf X}^{n+1})$, reject the move:  ${\bf X}^{n+1}
  = {\bf X}^{n}$. 
  \end{itemize}

  \end{itemize}
It is not difficult to show (see~\cite{MT} for instance) that, for a very
large class of transition densities $T$, the points ${\bf X}^n$
generated by the 
Metropolis algorithm are asymptotically distributed according to the
probability density $f({\bf X}) / \int_{\RR^d} f$. On the other hand,
the practical efficiency of the algorithm crucially depends on the
choice of the transition density (i.e. of the Markov chain).

\subsection{Random walks in the configuration space}  \label{sec:RW_CS}

In this section, the state space is the configuration space $\RR^{3N}$
and $f = |\Psi|^2$, so that the Metropolis algorithm actually
samples the probability density
$|\Psi({\bf R})|^2/\int_{\RR^{3N}} |\Psi|^2$.

\subsubsection{Simple random walk} 

In the original paper~\cite{metropolis} of Metropolis {\it et al.}, the
Markov chain is a simple random walk:
\begin{equation}
\widetilde{\bf R}^{n+1} = {\bf R}^n +  \Delta R \; {\bf U}^n
\end{equation}
where $\Delta R$ is the step size and ${\bf U}^n$ are independent and
identically distributed (i.i.d.) random vectors drawn uniformly in the
$3N$-dimensional cube $K = [-1,1]^{3N}$. The corresponding transition
density is $T({\bf R} \rightarrow {\bf R}') = \dps 
2^{-3N} \, \chi_{K}\left( ({\bf R}-{\bf R}')/\Delta R\right)$
where $\chi_K$ is 
the characteristic function of the cube $K$ (note that in this
particular case, $T({\bf R} \rightarrow {\bf R}') = T({\bf R}'
\rightarrow {\bf R})$). 

\subsubsection{Biased random walk}

The simple random walk is far from being the optimal choice: it induces a
high rejection rate, hence a large variance. A variance reduction
technique usually referred to as the importance sampling method, consists
in considering the so-called biased random walk or
over-damped Langevin dynamics~\cite{ceperley}: 
\begin{equation}
 \text{d}{\bf R}(t) = \nabla [ \log |\Psi| ] ({\bf R}(t))
 \text{d}t + \text{d}{\bf W}(t), 
 \label{eq:euler_cont}
\end{equation}
where ${\bf W}(t)$ is a $3N$-dimensional Wiener
process. Note that $|\Psi|^2$ is an invariant measure of the Markov
process~(\ref{eq:euler_cont}), and, better, that the
dynamics~(\ref{eq:euler_cont}) is in fact ergodic and satisfies a
detailed balance property~\cite{MT}. The qualifier ergodic means that
for any compactly supported continuous function $g \, : \, \RR^{3N}
\longrightarrow \RR$,
\begin{equation}
\lim_{T \rightarrow +\infty} \frac 1 T \int_0^T g( {\bf R}(t) ) \, dt
= \frac{\dps \int_{\RR^{3N}} g( {\bf R} ) \, |\Psi( {\bf R} )|^2 \, 
\text{d}{\bf R}}{\dps  \int_{\RR^{3N}} |\Psi(
{\bf R} )|^2 \, \text{d}{\bf R}}.
\end{equation}
The detailed balance property reads
\begin{equation}
|\Psi({\bf R})|^2  \, T_{\Delta t} ({\bf R} \rightarrow {\bf R}') \,
= \,
|\Psi({\bf R'})|^2 \,  T_{\Delta t} ({\bf R}' \rightarrow {\bf R})
\end{equation}
for any $\Delta t > 0$,
where $T_{\Delta t} ({\bf R} \rightarrow {\bf R}')$ is the
probability density that the Markov process~(\ref{eq:euler_cont}) is at 
${\bf R}'$ at time $t+\Delta t$ if it is at ${\bf R}$ at time
$t$. 
These above results are classical for regular,
positive functions $\Psi$, and have been recently proven for fermionic
wave functions~\cite{CJL} (in the latter case, the dynamics is ergodic in
each nodal pocket of the wave function $\Psi$).

Note that if one uses the Markov chain of
density $T_{\Delta t} ({\bf R} \rightarrow {\bf R}')$ in the
Metropolis algorithm, the acceptation/rejection step is useless, since
due to the detailed balance property, the acceptance rate always equals
one.

The exact value of $T_{\Delta t} ({\bf R} \rightarrow
{\bf R}')$ being not known,   
a discretization of equation~(\ref{eq:euler_cont}) with Euler scheme, is
generally used 
\begin{equation}
{\bf R}^{n+1} = {\bf R}^n + \Delta t \, \nabla [ \log |\Psi| ]
 ({\bf R}^n) + \Delta {\bf W}^n
 \label{eq:euler_disc}
\end{equation}
where $\Delta {\bf W}^n$ are
i.i.d. Gaussian random vectors with zero mean and covariance matrix 
$\Delta t \, I_{3N}$ ($I_{3N}$ is the identity matrix). The Euler scheme
leads to the approximated transition density
\begin{widetext}
\begin{equation}
T_{\Delta t}^{\rm Euler}({\bf R} \rightarrow {\bf R}')  
=  \frac{1}{(2\pi\Delta t)^{3N/2}} \; 
\exp \left( - \frac{\dps 
\left| {\bf R}' - {\bf R} - \Delta t \, \nabla [ \log
  |\Psi| ] ({\bf R}) \right|^2}{2\Delta t} \right)
\end{equation}
\end{widetext}
The time discretization introduces the so-called 
{\em time-step error}, whose consequence is that (\ref{eq:euler_disc})
samples $|\Psi({\bf R})|^2/\int_{\RR^{3N}} |\Psi|^2|$
{\em only approximately}. Note that the Metropolis 
acceptation/rejection procedure perfectly corrects the time-step error. 
In the limit $\Delta t \rightarrow 0$, the time-step error vanishes and
the acceptation/rejection procedure is useless.

This sampling method is much more efficient than the Metropolis
algorithm based on the simple random walk,
since the Markov chain (\ref{eq:euler_disc}) does a large part of the
work (it samples a short time-step approximation of $|\Psi({\bf R})|^2/\int
|\Psi^2|$), which is clearly not the case for the simple random walk.

The standard method in VMC computations currently is the Metropolis
algorithm based on the Markov chain defined by~(\ref{eq:euler_disc}).
For refinements of this method, we refer to~\cite{umrigar,sun,bressanini2}.

\subsection{Random walks in the phase space} \label{sec:RW_PS}

In this section, the state space is the phase space $\RR^{3N} \times
\RR^{3N}$. Let us emphasize that the introduction of momentum variables
in nothing but a numerical artifice. The phase space trajectories that
will be dealt with in this section do not have any physical meaning.

\subsubsection{Langevin dynamics}
\label{sec:langevin}

The Langevin dynamics of a system of $N$ particles of mass $m$ evolving
in an external potential $V$ reads 
\begin{equation}\label{eq:langevin}
\left\{
\begin{array}{lcl}
\text{d}{\bf R}(t) & = & \frac{1}{m} {\bf P}(t) \text{d}t, \\
\text{d}{\bf P}(t) & = & -\nabla V({\bf R}(t)) \text{d}t - \gamma {\bf P}(t)
\text{d}t + \sigma \text{d}{\bf W}(t).
\end{array}
\right.
\end{equation}
As above, ${\bf R}(t)$ is a $3N$-dimensional vector collecting the
positions at time $t$ of the $N$ particles. The components of the
$3N$-dimensional vector ${\bf P}(t)$ are the corresponding momenta
and ${\bf W}(t)$ is a $3N$-dimensional Wiener process.
The Langevin dynamics can be considered as a perturbation of the Newton
dynamics (for which $\gamma = 0$ and $\sigma = 0$). The magnitudes
$\sigma$ and $\gamma$ of the random forces $\sigma 
\text{d}{\bf W}(t)$ and of the drag term $- \gamma
{\bf P}(t)\text{d}t$ are related through the fluctuation-dissipation
formula 
\begin{equation}\label{eq:sigma}
\sigma^2 = \frac{2 m \gamma}{\beta},
\end{equation}
where $\beta$ is the reciprocal temperature of the system. Let us
underline that in the present setting, $\beta$ is a numerical parameter
that is by no means related to the physical temperature of the system. 
It can be checked (at least for regular potentials $V$) that the canonical
distribution   
\begin{equation}
 \text{d}\Pi ({\bf R},{\bf P}) = Z^{-1} e^{-\beta
  H({\bf R},{\bf P})} 
  \text{d}{\bf R} \text{d}{\bf P}
\end{equation}
is an invariant probability measure for the system, $Z$ being a
normalization constant and 
\begin{equation}
 H({\bf P},{\bf R}) = V({\bf R}) + \frac{|{\bf P}|^2}{2m}
\end{equation}
being the Hamiltonian of the underlying Newton dynamics. In addition,
the Langevin dynamics is ergodic (under some assumptions on $V$).
Thus, choosing
\begin{equation}
\text{$\beta=1$ and $V=-\log |\Psi|^2$,}
\end{equation}
the projection on the position space of the Langevin dynamics
samples $|\Psi({\bf R})|^2/\int |\Psi|^2$.  On the
other hand, the Langevin dynamics does not satisfy the detailed
balance property. We will come back to this important point in the
forthcoming section.

In this context, the parameters $m$ and $\gamma$ ($\sigma$ being
then obtained through~(\ref{eq:sigma})) should be seen as numerical
parameters to be optimized to get the best sampling. We now describe how
to discretize and apply a Metropolis algorithm
to the Langevin dynamics~(\ref{eq:langevin}), in the context of VMC.

\subsubsection{Time discretization of the Langevin dynamics}

Many discretization schemes exist for Langevin dynamics. In order to choose
which algorithm is best for VMC, we have tested four different schemes
available in the literature~\cite{bbk,allen,stoltz,ricci}, with parameters
$\beta=1$, $\gamma=1$ and $m=1$. Our benchmark system is a Lithium atom,
and a single determinantal wave function built upon Slater-type atomic
orbitals, multiplied by a Jastrow factor.
We turn off the acceptation/rejection step in these preliminary tests,
since our purpose is to compare the time-step errors for the various
algorithms. From the results displayed in 
table~\ref{tab:langevin}, one can see that the Ricci-Ciccotti
algorithm~\cite{ricci} is the method which generates the smallest time-step
error. This algorithm reads
\begin{widetext}
\begin{equation}\label{eq:ricci}
\left\{ 
\begin{array}{lcl}
{\bf R}^{n+1} & = & {\bf R}^n + \frac{\Delta t}m {\bf P}^n 
e^{-\gamma \Delta t/2} 
                       + \frac{\Delta t}{2m}  \big[ -\nabla V (
                       {\bf R}^n )\Delta t  + {\bf G}^n \big]
                       e^{-\gamma \Delta t/4}, \\ 
{\bf P}^{n+1} & = & {\bf P}^n e^{-\gamma \Delta t} - \frac{\Delta t}{2}
                     \big[ \nabla V ( {\bf R}^n ) 
                     + \nabla V ( {\bf R}^{n+1} ) \big]
                     e^{-\gamma \Delta t/2} + {\bf G}^n e^{-\gamma
                       \Delta t/2}, 
\end{array}
\right.
\end{equation}
\end{widetext}
where ${\bf G}^n$ are i.i.d. Gaussian random vectors with
zero mean and variance $\sigma^2 I_{3N}$ with $\sigma^2 = \frac{ 
  2 \gamma m}{\beta} \Delta t$.

It can be seen from Table I that the Ricci-Ciccotti algorithm also
outperforms the biased random walk~(\ref{eq:euler_disc}), as far as
sampling issues are concerned. In the following, we shall therefore use
the Ricci-Ciccotti algorithm.

\subsubsection{Metropolized Langevin dynamics}

The discretized Langevin dynamics does not exactly sample the
target distribution $\Pi$, but rather from some approximation
$\Pi_{\Delta t}$ of $\Pi$. It is therefore tempting to introduce a
Metropolis acceptation/rejection step to further improve the quality of
the sampling. Unfortunately, this idea cannot be straightforwardly
implemented for two reasons: 
\begin{itemize}
\item first, this is not technically feasible, for the Markov chain defined by
(\ref{eq:ricci}) does not have a transition density. Indeed, as the same 
Gaussian random vectors ${\bf G}^n$ are used to update both the positions
and the momenta, the measure $p(({\bf R}^n,{\bf P}^n),\cdot)$ is
supported on a  
$3N$-dimensional submanifold of the phase space $\RR^{3N} \times
\RR^{3N}$;
\item second, leaving apart the above mentioned technical difficulty,
  which is specific to the Ricci-Ciccotti scheme,
  the Langevin dynamics is {\it a priori} not an efficient Markov chain
  for the Metropolis algorithm, for it does not satisfy the detailed
  balance property.
\end{itemize}
Let us now explain how to tackle these two issues, starting with the
first one.

To make it compatible with the Metropolis framework, one needs
to slightly modify the Ricci-Ciccotti
algorithm. Following~\cite{C43,allen}, we thus introduce   
i.i.d. {\em correlated} Gaussian vectors $(G_{1,i}^n,G_{2,i}^n)$ ($1 \le
i \le 3N$) such that: 
\begin{subequations}
\begin{align}
\langle(G_{1,i}^n)^2\rangle= \sigma_{1}^2 & = \frac{\Delta t}{\beta m
   \gamma} \left( 2 - 
   \frac{3 - 4 e^{-\gamma \Delta t} + e^{-2 \gamma \Delta t}}{\gamma \Delta t}
   \right), \\
 \langle(G_{2,i}^n)^2\rangle=\sigma_{2}^2 & = \frac{m}{\beta} \left( 1-e^{-2 \gamma \Delta t} \right), \\
 \frac{\langle G_{1,i}^n G_{2,i}^n \rangle}{\sigma_{1}\sigma_{2}}=c_{12} & = \frac{(1-e^{-\gamma \Delta t} )^2}{\beta \gamma \sigma_{1}\sigma_{2}}.
\end{align}
\end{subequations}
Setting ${\bf G}^n_1 = (G_{1,i}^n)_{1 \le i \le 3N}$ and 
${\bf G}^n_2 = (G_{2,i}^n)_{1 \le i \le 3N}$, the modified
Ricci-Ciccotti algorithm reads
\begin{widetext}
\begin{equation}\label{eq:ricci_metropolis}
\left\{ 
\begin{array}{lcl}
{\bf R}^{n+1} & = & {\bf R}^n + \frac{\Delta t}m {\bf P}^n 
e^{-\gamma \Delta t/2} 
                       - \frac{\Delta t^2}{2m}  \nabla V ( {\bf R}^n )
                       e^{-\gamma \Delta t/4} + {\bf G}_{1}^n,\\
{\bf P}^{n+1} & = & {\bf P}^n e^{-\gamma \Delta t} - \frac{\Delta t}{2}
                     \big[ \nabla V ( {\bf R}^n ) 
                       + \nabla V ( {\bf R}^{n+1} ) \big]
                     e^{-\gamma \Delta t/2} + {\bf G}_{2}^n.
\end{array}
\right.
\end{equation}
The above scheme is a consistent discretization of (\ref{eq:langevin})
and the corresponding Markov chain does have a transition density, which
reads (see Appendix) 
\begin{subequations}\label{eq:T}
\begin{align}
T^{\rm MRC}_{\Delta t}(({\bf R}^n, {\bf P}^n) & \rightarrow
({\bf R}^{n+1}, 
{\bf P}^{n+1})) = \nonumber \\ 
&Z^{-1} \exp \left[ -\frac{1}{2(1-c_{12}^2)}
\left( \left( \frac{|{\bf d}_1|}{\sigma_{1}} \right)^2 +
\left( \frac{|{\bf d}_2|}{\sigma_{2}} \right)^2  - 2 c_{12}
\frac{{\bf d}_1}{\sigma_{1}} \cdot
\frac{{\bf d}_2}{\sigma_{2}} \right)
\right], \label{eq:Ta}
\end{align}
\text{with}
\begin{align}
{\bf d}_1 &= {\bf R}^{n+1} - {\bf R}^n - \Delta t
\frac{{\bf P}^n}{m} 
 e^{-\gamma \Delta t/2} + \frac{\Delta t^2}{2m} \nabla V ( {\bf R}^n )
 e^{-\gamma \Delta t/4}, \label{eq:Tb}\\
{\bf d}_2 &= {\bf P}^{n+1} - {\bf P}^n e^{-\gamma \Delta t}
 + \frac{1}{2} \Delta t  \big[ \nabla V ( {\bf R}^n ) + \nabla V (
 {\bf R}^{n+1} ) 
 \big] e^{-\gamma \Delta t/2}.\label{eq:Tc}
\end{align}
\end{subequations} 
\end{widetext}
Unfortunately, inserting directly the transition density (\ref{eq:T}) in the
Metropolis algorithm leads to a high rejection rate. Indeed, if
$({\bf R}^n,{\bf P}^n)$ and $({\bf R}^{n+1}, {\bf P}^{n+1})$
are related through~(\ref{eq:ricci_metropolis}), 
$T^{\rm MRC}_{\Delta t}(({\bf R}^n,{\bf P}^n) \rightarrow
({\bf R}^{n+1},{\bf P}^{n+1}))$ usually is much greater than 
$T^{\rm MRC}_{\Delta t}(({\bf R}^{n+1}, {\bf P}^{n+1} ) \rightarrow
({\bf R}^n, {\bf P}^n))$,
 since the probability that the random forces are strong enough to make
 the particle go back in one step from where it comes, is very low in
 general. This is related to the fact that the Langevin dynamics does
 not satisfy the detailed balance relation.

It is however possible to further modify the overall algorithm by ensuring
some microscopic reversibility, in order
to finally obtain low rejection rates. For this purpose, we
introduce momentum reversions. Denoting by $T^{\rm Langevin}_{\Delta t}$
the transition density of the 
Markov chain obtained by  integrating~(\ref{eq:langevin})
{\em exactly} on the time interval $[t,t+\Delta t]$, it is indeed not
difficult to 
check (under convenient assumptions on $V= - \log |\Psi|^2$, that the
Markov chain defined by the transition density  
\begin{widetext}
\begin{equation}
\widetilde T^{\rm Langevin}_{\Delta t} 
\left(({\bf R},{\bf P}) \rightarrow ({\bf R}',{\bf P}')
\right)  = T^{\rm Langevin}_{\Delta t} 
\left(({\bf R},{\bf P}) \rightarrow ({\bf R}',-{\bf P}')
\right)
\end{equation}
is ergodic with respect to $\Pi$ and satisfies the detailed balance
property 
\begin{equation}
 \Pi({\bf R},{\bf P}) \, \widetilde T^{\rm Langevin}_{\Delta t} 
\left(({\bf R},{\bf P}) \rightarrow ({\bf R}',{\bf P}')
\right)  = \Pi({\bf R}',{\bf P}') \,
\widetilde T^{\rm Langevin}_{\Delta t} 
\left(({\bf R}',{\bf P}') \rightarrow ({\bf R},{\bf P})
\right)  .
\end{equation}
\end{widetext}
Replacing the exact transition density  $T^{\rm Langevin}_{\Delta t}$ by
the approximation $T^{\rm MRC}_{\Delta t}$, we now consider the
transition density
\begin{widetext}
\begin{equation}
\widetilde T^{\rm MRC}_{\Delta t} \left(({\bf R},{\bf P})
  \rightarrow ({\bf R}',{\bf P}')^{\rm MRC}_{\Delta t} 
\right)  = T^{\rm MRC}_{\Delta t} 
\left(({\bf R},{\bf P}) \rightarrow ({\bf R}',-{\bf P}')
\right).
\end{equation}
\end{widetext}
The new sampling algorithm that we propose can be stated as follows: 
\begin{itemize}
\item Propose a move from $({\bf R}^n,{\bf P}^n)$ to
$(\widetilde{{\bf R}}^{n+1},\widetilde{{\bf P}}^{n+1})$
using the transition density~$\widetilde T^{\rm MRC}_{\Delta t}$. In
other words, perform one step of the modified Ricci-Ciccotti algorithm
\begin{widetext}
\begin{equation}
\left\{ 
\begin{array}{lcl}
{\bf R}^{n+1}_\ast & = & {\bf R}^n + \frac{\Delta t}m {\bf P}^n 
e^{-\gamma \Delta t/2} 
                       - \frac{\Delta t^2}{2m}  \nabla V ( {\bf R}^n )
                       e^{-\gamma \Delta t/4} + {\bf G}_{1}^n,\\
{\bf P}^{n+1}_\ast & = & {\bf P}^n e^{-\gamma \Delta t} -
\frac{\Delta t}{2} 
                     \big[ \nabla V ( {\bf R}^n ) 
                       + \nabla V ( {\bf R}^{n+1} ) \big]
                     e^{-\gamma \Delta t/2} + {\bf G}_{2}^n.
\end{array}
\right.
\end{equation}
\end{widetext}
and set $(\widetilde{{\bf R}}^{n+1},\widetilde{{\bf P}}^{n+1}) = 
({\bf R}^{n+1}_\ast,-{\bf P}^{n+1}_\ast)$

\item Compute the acceptance rate
\begin{widetext}
\begin{eqnarray*}
\lefteqn{{A}(({\bf R}^n, {\bf P}^{n}) 
\rightarrow (\widetilde{{\bf R}}^{n+1}, \widetilde{{\bf P}}^{n+1}))}\\
& \displaystyle{= \min \left( \frac{\Pi
      ({\bf R}^{n+1},{\bf P}^{n+1})  
\, \widetilde T^{\rm MRC}_{\Delta t}  ((\widetilde{{\bf R}}^{n+1}, \widetilde{{\bf P}}^{n+1})  
\rightarrow({\bf R}^n,  {\bf P}^{n})  )
}
     {\Pi({\bf R}^{n},{\bf P}^{n})  \, \widetilde T^{\rm
         MRC}_{\Delta t}(({\bf R}^n, 
       {\bf P}^{n})  
\rightarrow (\widetilde{{\bf R}}^{n+1}, \widetilde{{\bf P}}^{n+1}))
},\,1\right).} 
\end{eqnarray*}
\end{widetext}
\item Draw a random variable $U^n$ uniformly distributed in $(0,1)$ and
  \begin{itemize}
  \item if $U^n \leq  {A}(({\bf R}^n, {\bf P}^{n}) 
\rightarrow (\widetilde{{\bf R}}^{n+1},
\widetilde{{\bf P}}^{n+1}))$, accept the proposal:  
$(\overline{{\bf R}}^{n+1},\overline{{\bf P}}^{n+1}) = (\widetilde{{\bf R}}^{n+1},
\widetilde{{\bf P}}^{n+1})$,
\item if $U^n > {A}(({\bf R}^n, {\bf P}^{n}) 
\rightarrow (\widetilde{{\bf R}}^{n+1},
\widetilde{{\bf P}}^{n+1}))$, reject the proposal, and set 
$(\overline{{\bf R}}^{n+1},\overline{{\bf P}}^{n+1}) =
({\bf R}^{n}, {\bf P}^{n})$. 
  \end{itemize}
\item Reverse the momenta
\begin{equation}
({\bf R}^{n+1}, {\bf P}^{n+1}) =
(\overline{{\bf R}}^{n+1},- \overline{{\bf P}}^{n+1}) 
\end{equation}
\end{itemize}
Note that a momentum reversion is systematically performed just after
the Metropolis step. As the invariant measure $\Pi$ is left unchanged by
this operation, the global algorithm (Metropolis step based on the
transition density $\widetilde T^{\rm MRC}_{\Delta t}$ plus momentum
reversion) actually samples $\Pi$. The role of the final momentum
reversion is to preserve the underlying Langevin dynamics: while the
proposals are accepted, the above algorithm generates Langevin
trajectories, that are known to efficiently sample an approximation 
of the target density $\Pi$. Numerical tests seem to show that, in
addition, the momentum reversion also plays a role when the proposal is
rejected: it seems to increase the acceptance rate of the next step,
preventing the walkers from being trapped in the vicinity of the
nodal surface $\Psi^{-1}(0)$. 

\medskip

\noindent
As the points $({\bf R}^n,{\bf P}^n)$ of the phase space generated
by the above algorithm form a sampling of $\Pi$, the positions
$({\bf R}^n)$  sample $|\Psi({\bf R})|^2/ \int_{\RR^{3N}}|\Psi|^2$
and can therefore be used for VMC calculations.

\section{Numerical experiments and applications} \label{sec:num}

\subsection{Measuring the efficiency}
\label{sec:proba}

A major drawback of samplers based on Markov processes is that they
generate sequentially correlated data. The effective number of
independent observations is in fact $L_\text{eff} = 
L/N_\text{corr}$, where $N_\text{corr}$ is the {\em correlation
  length}, namely the number of successive correlated moves.

In the following applications, we provide estimators for the
correlation length $N_\text{corr}$ and for the so-called inefficiency
$\eta$ (see below), which are relevant indicators of the quality of the
sampling. In this section, following Stedman \emph{et al.}~\cite{stedman},
we describe the way these quantities are defined and computed.

The sequence of samples is split into $N_B$ blocks of $L_B$ steps, where the
number $L_B$ is chosen such that it is a few orders of magnitude higher than
$N_\text{corr}$.
The empirical mean of the local energy reads
\begin{equation}
 \langle E_{L} \rangle_{|\Psi|^2}^{N_B,L_B} = \frac{1}{N_B L_B}
 \sum_{i=1}^{N_B L_B} E_{L}({\bf R}^i).
\end{equation}
The empirical variance over all the individual steps is given by
\begin{equation}
 [\sigma^{N_B,L_B}]^2 = 
\frac{1}{N_B L_B} \sum_{i=1}^{N_B L_B} \left(
 E_{L}({\bf R}^i) - \langle E_{L}
 \rangle_{|\Psi|^2} \right)^2 
\end{equation}
and the empirical variance over the blocks by
\begin{equation}
 [\sigma_B^{N_B,L_B}]^2  = \frac{1}{N_B} \sum_{i=1}^{N_B} \left(
 E_{B,i} - \langle E_{L} \rangle_{|\Psi|^2}^{N_B,L_B} \right)^2,
\end{equation}
where $E_{B,i}$ is the average energy over block $i$:
\begin{equation}
 E_{B,i} = \frac{1}{L_B} \sum_{j=(i-1)L_B+1}^{i L_B}
 E_{L}({\bf R}^j).
\end{equation}

Following~\cite{stedman}, we define the correlation length as
\begin{equation}\label{eq:N_corr}
 N_\text{corr} = \lim_{N_B\rightarrow \infty}  \lim_{L_B\rightarrow \infty} L_B \frac{[\sigma_B^{N_B,L_B}]^2}{[\sigma^{N_B,L_B}]^2},
\end{equation}
and the inefficiency $\eta$ of the run as:
\begin{equation}
 \eta =\lim_{N_B\rightarrow \infty} \lim_{L_B\rightarrow \infty}  L_B
 [\sigma_B^{N_B,L_B}]^2. 
\end{equation}
On the numerical examples presented below, the relative fluctuations of the
quantities $ L_B \frac{[\sigma_B^{N_B,L_B}]^2}{[\sigma^{N_B,L_B}]^2}$ and $L_B
 [\sigma_B^{N_B,L_B}]^2$ become small for $L_B > 50$ and $N_B > 50$.

The definition of these two quantities can be understood as follows. Since $L_B \gg N_\text{corr}$
and only $L_B/N_\text{corr}$ are independent samples among the samples
in the block, the central limit theorem yields $E_{B,i} \simeq \langle E_{L} \rangle_{|\Psi|^2} +
\frac{\sigma \, G^i}{\sqrt{L_B/ N_\text{corr}}}$ where $G^i$ are
i.i.d. normal random variables. Thus, in the limit $N_B\rightarrow
\infty$ and $L_B\rightarrow \infty$, we obtain that
$\sigma_B^2=\frac{\sigma^2}{L_B/ N_\text{corr}}$ which
yields~(\ref{eq:N_corr}). The inefficiency $\eta$ is thus equal to
$N_\text{corr} \sigma^2$ and is large if the variance is large, or if
the number of correlated steps is large.

Using this measure of efficiency, we can now compare the
sampling algorithms (the simple random walk, the
biased random walk and 
the Langevin algorithm) for various systems.  In any case, a
Metropolis 
acceptation/rejection step is used. We found empirically from several
tests that convenient values for the
parameters of the Langevin algorithm are $\gamma=1$ and $m=Z^{3/2}$ where
$Z$ is the highest nuclear charge among all the nuclei. For each
algorithm, we compare the efficiency for various values of the step
length, namely the increment $\Delta R$ in the case of the simple random
walk, and the time-step $\Delta t$ for
the other two schemes.  For a given algorithm, simple arguments
corroborated by numerical tests show that there exists an optimal
value of this increment: for smaller (resp. for larger increments), the
correlation between two successive positions increases since the displacement
of the particle is small (resp. since many moves are rejected), and this
increases the number of correlated steps $N_\text{corr}$.

One can notice on the results (see
tables~\ref{tab:lithium},~\ref{tab:fluorine_sto},~\ref{tab:copper},~\ref{tab:phenol})
that a large error bar corresponds to large values for $N_\text{corr}$
and $\eta$. The quantities $N_\text{corr}$
and $\eta$ are a way to refine the measure of efficiency, since the same
length of error bar may be obtained for different values of the numerical parameters.

Let us now present some numerical tests. We compare the
algorithms and parameters at a fixed computational cost. The reference
values are 
obtained by ten times longer VMC simulations. The error bars given in
parenthesis are $60\%$ confidence intervals. We also provide the
acceptance rate (denoted by $A$ in the tables) and, when it is
relevant, the mean of
the length of the increment ${\bf R}^{n+1}-{\bf R}^n$ over one
time-step (denoted by $\langle |\Delta 
{\bf R}| \rangle$ in the tables) for the biased random walk and the
Langevin dynamics.

\subsection{Atoms}

\emph{Lithium}.
The Lithium atom was chosen as a first simple example. The
wave function is the same as for the benchmark system used for the
comparison of the 
various Langevin schemes, namely
a single Slater determinant of Slater-type basis 
functions improved by a Jastrow factor to take account of the electron
correlation. The reference energy associated
with this 
wave function is $-7.47198(4)$ a.u., and the comparison of the algorithms is
given in table~\ref{tab:lithium}.
The runs were made of 100 random walks
composed of 50 blocks of 1000 steps.
For the simple random walk, the lowest values of the correlation length
and of the inefficiency are respectively $11.4$ and $1.40$. The biased
random walk is
much more efficient, since the optimal correlation length and inefficiency are
more than twice smaller, i.e. $4.74$ and $0.55$.
The proposed algorithm is even more efficient: the optimal correlation length is
$3.75$ and the optimal inefficiency is $0.44$.

\emph{Fluorine}.
The Fluorine atom was chosen for its relatively ``high'' nuclear charge
($Z=9$), leading to a timescale separation of the core and valence
electrons.
The wave function is a Slater-determinant with Gaussian-type basis functions
where the $1s$ orbital was substituted by a Slater-type orbital, with a
reference energy of $-99.397(2)$ a.u.
The runs were made of 100 random walks
composed of 100 blocks of 100 steps.
The results are given in table~\ref{tab:fluorine_sto}.
For the simple random walk, the lowest values of the correlation length
and of the inefficiency are respectively $15.6$ and $282$. The biased
random walk is
again twice more efficient than the simple random walk, for which the optimal
correlation length and inefficiency are $7.4$ and $137$.
The Langevin algorithm is more efficient than the biased random walk:
the optimal correlation length is $5.3$ and the optimal inefficiency is $102$.

\emph{Copper}.
We can go even further in the timescale separation and take the Copper atom
($Z=29$) as an example. The wave function is a Slater determinant with a basis
of Slater-type atomic orbitals, improved by a Jastrow factor to take account of
the electron correlation. The reference energy is $-1639.2539(24)$.
The runs were made of 40 random walks
composed of 500 blocks of 500 steps. From table~\ref{tab:copper}, 
one can remark that the Langevin algorithm is
again more efficient than the biased random walk, since the optimal correlation
length and inefficiency are respectively $28.7$ and $4027$, whereas using the
biased random walk, these values are $51.0$ and $5953$.

\subsection{The phenol molecule}

The Phenol molecule was chosen to test the proposed algorithm because it
contains three different types of atoms (H, C and O). The wave function here is
a single Slater determinant with Gaussian-type basis functions. The core
molecular orbitals of the Oxygen and Carbon atoms were substituted by the
corresponding atomic $1s$ orbitals. The comparison of the biased random walk
with the Langevin
algorithm is given in table~\ref{tab:phenol}.
The optimal correlation length using the biased random walk is 10.17, whereas it is
8.23 with our Langevin algorithm. The optimal inefficiency is again lower with the
Langevin algorithm (544) than with the biased random walk (653).

\subsection{Discussion of the results}

We observe that on our numerical tests, the Langevin dynamics is always more
efficient than the biased
random walk. Indeed, we notice that:
\begin{itemize}
\item The error bar (or $N_{\text{corr}}$, or $\eta$) obtained with the Langevin dynamics for an optimal
  set of numerical parameters is always smaller than the error bar
  obtained with other algorithms (for which we also optimize the numerical parameters).
\item The size of the error bar does not seem to be as sensitive to the
  choice of the numerical parameters as for other methods. In
  particular, we observe on our numerical tests that the value $\Delta
  t=0.2$ seems to be convenient to obtain good results with the Langevin
  dynamics, whatever the atom or molecule.
\end{itemize}

\section*{Acknowledgments} This work was supported by the ACI
``Molecular Simulation'' of the French Ministry of Research.

\appendix

\section{Derivation of the transition probability~(\ref{eq:T})}

The random vector
$({\bf d}_1,{\bf d}_2)$ (defined by~(\ref{eq:Tb})--(\ref{eq:Tc}))
is a Gaussian random vector and therefore admits a density with respect
to the Lebesgue measure in $\mathbb{R}^{6N}$. If, for $1 \leq i \leq 3N$, we denote by $d_{1,i}$
(resp. $d_{2,i}$) the components of ${\bf d}_1$
(resp. ${\bf d}_2$), we observe that the Gaussian random vectors
$(d_{1,i},d_{2,i})$ are i.i.d. Therefore, the transition probability
$T(({\bf R}^n, {\bf P}^n) \rightarrow ({\bf R}^{n+1}, 
{\bf P}^{n+1}))$ reads
\begin{equation}\label{eq:1}
T(({\bf R}^n, {\bf P}^n) \rightarrow ({\bf R}^{n+1}, 
{\bf P}^{n+1})) =Z^{-1} \left(p(d_{1,i},d_{2,i})\right)^{3N}
\end{equation}
where $Z$ is a normalization constant and $p$ denotes the density (in $\mathbb{R}^2$) of the Gaussian random vectors
$(d_{1,i},d_{2,i})$.

From equations~(\ref{eq:ricci_metropolis}), one can see that
\begin{widetext}
\begin{eqnarray*}
d_{1,i} &=& R^{n+1}_i - R^n_i - \Delta t \frac{P^n_i}{m}
 e^{-\gamma \Delta t/2} + \frac{\Delta t^2}{2m} \nabla_i V ( {\bf R}^n )
 e^{-\gamma \Delta t/4}, \\
d_{2,i} &=& P^{n+1}_i - P^n_i e^{-\gamma \Delta t}
 + \frac{1}{2} \Delta t  \big[ \nabla_i V ( {\bf R}^n ) + \nabla_i V ( {\bf R}^{n+1} )
 \big] e^{-\gamma \Delta t/2},
\end{eqnarray*}
\end{widetext}
is a Gaussian random vector with covariance matrix $\Gamma=\left[
\begin{array}{cc}
\sigma_1^2 & c_{12}\sigma_1\sigma_2 \\
c_{12}\sigma_1\sigma_2 & \sigma_2^2 \\
\end{array}
\right]$. Thus
\begin{equation}\label{eq:2}
p(d_1,d_2)=\left(2 \pi \sqrt{\det
   \Gamma}\right)^{-1} \exp \left(- \frac{1}{2} (d_1,d_2) \Gamma^{-1}
(d_1,d_2)^T\right).
\end{equation}
Since $\Gamma^{-1}=\frac{1}{(1-c_{12}^2)}\left[\begin{array}{cc}
\frac{1}{\sigma_1^2} & -\frac{c_{12}}{\sigma_1\sigma_2} \\
-\frac{c_{12}}{\sigma_1\sigma_2} & \frac{1}{\sigma_2^2} \\
\end{array}\right]$, (\ref{eq:Tc}) is easily obtained from~(\ref{eq:1})--(\ref{eq:2}).

%%%% TABLES
\newpage
\section*{Table captions}

\begin{itemize}
\item Table~\ref{tab:langevin}: Comparison of different discretization schemes
for Langevin dynamics. The reference energy is -7.47198(4) a.u.
\item Table~\ref{tab:lithium} : The Lithium atom: Comparison of the Simple random walk, the Biased random walk and the proposed Langevin algorithm. The runs were carried
out with 100 walkers, each realizing 50 blocks of 1000 steps. The
reference energy is -7.47198(4) a.u.
\item Table~\ref{tab:fluorine_sto} : The Fluorine atom : Comparison of
the Simple random walk, the Biased random walk and the proposed
Langevin algorithm. The runs were carried out with 100 walkers,
each realizing 100 blocks of 100 steps. The reference energy is
-99.397(2) a.u.
\item Table~\ref{tab:copper} : The Copper atom:
Comparison of the Biased random walk with the proposed Langevin algorithm.
The runs were carried out with 40 walkers, each realizing 500 blocks
of 500 steps. The reference energy is -1639.2539(24) a.u. 
\item Table~\ref{tab:phenol} : The Phenol molecule : Comparison of the
Biased random walk with the
proposed Langevin algorithm. The runs were carried out with 100 walkers,
each realizing 100 blocks of 100 steps. The reference energy is -305.647(2)
a.u.
\end{itemize}

%\newpage
\section*{Tables}
\begin{widetext}
\newpage

\begin{table}[h!]
 \begin{tabular}{lcccccc}
  \hline
  \hline
$\Delta t$
& BRW
& BBK~\cite{bbk}
& Force interpolation~\cite{allen} 
& Splitting~\cite{stoltz} 
& Ricci \& Ciccotti~\cite{ricci} \\
  \hline
0.05    & -7.3758(316)    & -7.4395(246)    & -7.4386(188)    & -7.4467(137)    & -7.4576(07) \\
0.005    & -7.4644(069)    & -7.4698(015)    & -7.4723(015)    & -7.4723(015)    & -7.4701(20)  \\
0.001    & -7.4740(007)    & -7.4728(013)    & -7.4708(017)    & -7.4708(017)    & -7.4696(17) \\
0.0005    & -7.4732(010)    & -7.4700(023)    & -7.4709(022)    & -7.4708(022)    & -7.4755(26) \\
  \hline
  \hline
 \end{tabular}
 \caption{Comparison of Langevin algorithms and the biased random walk (BRW)}
 \label{tab:langevin}
\end{table}
\newpage

\begin{table}[h!]
 \begin{tabular}{cccccc}
  \hline
  \hline
$\Delta R$    & $\langle E_L \rangle$    &$N_\text{corr}$    & $\eta$    & $A$ &\\
  \hline
{\em Simple random walk}&            &            &            &       \\
0.05    & -7.47126(183)    & 94.5 $\pm$ 3.3    & 11.72(42)    & 0.91 \\ 
0.10    & -7.47239(97)    & 35.2 $\pm$ 1.2    & 4.08(14)    & 0.82 \\ 
0.15    & -7.47189(75)    & 20.5(5)    & 2.30(06)    & 0.74 \\ 
0.20    & -7.47157(56)    & 14.3(4)    & 1.62(04)    & 0.66 \\ 
0.25    & -7.47182(56)    & 12.1(3)    & 1.40(05)    & 0.59 \\ 
0.30    & -7.47189(56)    & 11.4(3)    & 1.57(17)    & 0.52 \\ 
0.35    & -7.47275(59)    & 12.4(3)    & 1.57(17)    & 0.46 \\ 
0.40    & -7.47130(63)    & 14.4(5)    & 1.93(22)    & 0.40 \\ 
  \hline
$\Delta t$ & $\langle E_L \rangle$ & $N_\text{corr}$ & $\eta$ & $\langle |\Delta {\bf R}| \rangle$ & $A$ \\
  \hline
{\em Biased random walk}&            &            &            &       \\
0.01    & -7.47198(53)    & 10.31(29)    & 1.23(3)    & 0.284(09) & 0.98 \\ 
0.03    & -7.47156(39)    &  5.26(14)    & 0.73(7)    & 0.444(21) & 0.92 \\ 
0.04    & -7.47195(35)    &  4.82(12)    & 0.57(3)    & 0.486(26) & 0.88 \\ 
0.05    & -7.47219(32)    &  4.74(11)    & 0.55(2)    & 0.514(31) & 0.85 \\ 
0.06    & -7.47204(38)    &  4.95(11)    & 0.58(3)    & 0.533(36) & 0.81 \\ 
0.07    & -7.47251(32)    &  5.39(14)    & 0.61(3)    & 0.546(40) & 0.78 \\ 
0.10    & -7.47249(42)    &  7.56(25)    & 0.87(5)    & 0.555(50) & 0.68 \\ 
  \hline
{\em Langevin}&            &            &            &       \\
0.20    & -7.47233(34)    & 5.07(10)    & 0.60(1)    & 0.236(08) & 0.97 \\
0.30    & -7.47207(34)    & 4.14(09)    & 0.47(1)    & 0.328(15) & 0.93 \\
0.35    & -7.47180(31)    & 3.96(08)    & 0.45(1)    & 0.366(18) & 0.91 \\
0.40    & -7.47185(29)    & 3.75(08)    & 0.44(2)    & 0.399(22) & 0.89 \\
0.45    & -7.47264(29)    & 3.88(08)    & 0.45(2)    & 0.426(25) & 0.86 \\
0.50    & -7.47191(29)    & 4.07(14)    & 0.46(2)    & 0.426(25) & 0.84 \\
0.60    & -7.47258(32)    & 4.78(16)    & 0.52(2)    & 0.481(36) & 0.78 \\
  \hline
  \hline
 \end{tabular}
 \caption{The Lithium atom}
 \label{tab:lithium}
\end{table}
 
\newpage
\begin{table}[h!]
 \begin{tabular}{cccccc}
  \hline
  \hline
$\Delta R$    & $\langle E_L \rangle$    &$N_\text{corr}$    & $\eta$    & $A$ \\
  \hline
{\em Simple random walk}&            &            &            &       \\
0.02    & -99.398(72)    & 38.9(7)    & 823(31)    & 0.87 \\
0.05    & -99.426(39)    & 20.3(4)    & 405(11)    & 0.69 \\
0.08    & -99.406(28)    & 15.6(4)    & 326(17)    & 0.53 \\
0.10    & -99.437(23)    & 15.8(3)    & 282(07)    & 0.44 \\
0.12    & -99.402(24)    & 16.6(4)    & 341(24)    & 0.36 \\
0.15    & -99.398(25)    & 19.4(5)    & 412(41)    & 0.27 \\
  \hline
$\Delta t$ & $\langle E_L \rangle$ & $N_\text{corr}$ & $\eta$ & $\langle |\Delta {\bf R}| \rangle$ & $A$ \\
  \hline
{\em Biased random walk}    &        &            &            \\
0.002    & -99.411(21)    &  9.9(2)    & 206(04)    & 0.211(08)     & 0.94 \\
0.003    & -99.424(17)    &  8.8(2)    & 173(04)    & 0.242(11)     & 0.90 \\
0.004    & -99.430(15)    &  7.6(2)     & 147(03)    & 0.263(16)    & 0.86 \\
0.005    & -99.399(14)    &  7.3(2)     & 142(03)    & 0.275(17)    & 0.82 \\
0.006    & -99.406(14)    &  7.4(1)     & 137(03)    & 0.282(19)    & 0.79 \\
0.007    & -99.430(14)    &  7.4(2)     & 142(08)    & 0.286(21)    & 0.75 \\
0.008    & -99.421(13)    &  7.6(2)     & 141(05)    & 0.287(23)    & 0.71 \\
0.009    & -99.406(13)    &  7.8(2)     & 177(19)    & 0.285(25)    & 0.67 \\
0.010    & -99.419(15)    &  7.8(2)     & 162(10)    & 0.281(27)    & 0.64 \\
0.011    & -99.416(14)    &  8.3(2)     & 147(05)    & 0.276(28)    & 0.60 \\
0.012    & -99.420(15)    &  9.1(3)     & 205(34)    & 0.270(29)    & 0.57 \\
0.013    & -99.425(17)    & 10.2(4)     & 224(38)    & 0.263(30)    & 0.54 \\
  \hline
{\em Langevin}&        &            &            \\
0.10    & -99.402(16)    &  8.9(2)    & 199(04)    & 0.095(02)    & 0.98 \\
0.20    & -99.403(12)    &  6.0(1)    & 123(02)    & 0.174(06)    & 0.94 \\
0.25    & -99.402(12)    &  5.4(1)    & 108(02)    & 0.204(09)    & 0.91 \\
0.30    & -99.395(11)    &  5.3(1)    & 104(02)    & 0.228(10)    & 0.87 \\
0.35    & -99.409(12)    &  5.4(1)    & 108(06)    & 0.245(15)    & 0.83 \\
0.40    & -99.402(11)    &  5.5(1)    & 102(03)    & 0.256(18)    & 0.78 \\
0.45    & -99.406(11)    &  5.9(1)    & 114(06)    & 0.261(21)    & 0.73 \\
0.50    & -99.408(12)    &  6.6(2)    & 124(07)    & 0.262(24)    & 0.68 \\
0.55    & -99.407(14)    &  7.9(4)    & 149(10)    & 0.257(26)    & 0.62 \\
0.60    & -99.405(15)    &  9.2(4)    & 178(13)    & 0.250(42)    & 0.56 \\
  \hline
  \hline
 \end{tabular}
 \caption{The Fluorine atom}
 \label{tab:fluorine_sto}
\end{table}

\newpage
\begin{table}[h!]
 \begin{tabular}{cccccc}
  \hline
  \hline
$\Delta t$ & $\langle E_L \rangle$ & $N_\text{corr}$ & $\eta$ & $\langle |\Delta {\bf R}| \rangle$ & $A$ \\
  \hline
{\em Biased random walk}&            &            &            &       \\
0.0003  & -1639.2679( 78) &  79.1 $\pm$ 2.7 & 10682(420) & 0.1311(108) & 0.86 \\
0.0004  & -1639.2681( 98) &  70.4 $\pm$ 1.3 &  8682(204) & 0.1385(137) & 0.81 \\
0.0005  & -1639.2499( 96) &  61.3 $\pm$ 2.5 &  7770(297) & 0.1414(162) & 0.75 \\
0.0006  & -1639.2629( 96) &  56.0 $\pm$ 1.2 &  6834( 88) & 0.1414(183) & 0.70 \\
0.0007  & -1639.2575( 73) &  53.8 $\pm$ 0.8 &  6420( 81) & 0.1393(201) & 0.65 \\
0.00075 & -1639.2518( 85) &  53.1 $\pm$ 0.9 &  6330( 91) & 0.1377(209) & 0.62 \\
0.0008  & -1639.2370( 86) &  55.7 $\pm$ 3.6 &  6612(405) & 0.1357(216) & 0.60 \\
0.00105 & -1639.2694( 85) &  51.0 $\pm$ 0.8 &  5953( 90) & 0.1228(241) & 0.48 \\
0.0011  & -1639.2563(110) &  54.3 $\pm$ 1.8 &  6513(221) & 0.1198(245) & 0.46 \\
0.0012  & -1639.2523( 72) &  59.9 $\pm$ 5.5 &  7266(658) & 0.1136(251) & 0.43 \\
\hline
{\em Langevin}&        &  &            &            \\
0.05    & -1639.2553( 92) &  61.3 $\pm$ 1.7 & 8256(  89) & 0.0371(  1) & 0.99 \\
0.10    & -1639.2583( 76) &  40.6 $\pm$ 3.1 & 5319( 383) & 0.0705( 30) & 0.97 \\
0.15    & -1639.2496( 65) &  30.1 $\pm$ 0.8 & 4042( 103) & 0.0978( 60) & 0.93 \\
0.20    & -1639.2521( 71) &  28.7 $\pm$ 0.9 & 4027( 403) & 0.1173( 96) & 0.87 \\
0.30    & -1639.2510( 67) &  35.2 $\pm$ 2.5 & 4157( 291) & 0.1326(170) & 0.71 \\
0.40    & -1639.2524( 78) &  50.5 $\pm$ 3.7 & 5922( 455) & 0.1210(225) & 0.52 \\
\hline 
  \hline
 \end{tabular}
 \caption{The Copper atom}
 \label{tab:copper}
\end{table}

%\newpage
\begin{table}[h!]
 \begin{tabular}{cccccc}
  \hline
  \hline
$\Delta t$    & $\langle E_L \rangle$    &$N_\text{corr}$    & $\eta$    &
$\langle |\Delta {\bf R}| \rangle$    &    $A$ \\
  \hline
{\em Biased random walk}&            &            &            &            &       \\
0.003    & -305.6308(83)    & 18.71(24)    & 1368(12)    & 0.522(29)    & 0.85 \\
0.004    & -305.6471(78)    & 16.00(28)    & 1193(30)    & 0.547(36)    & 0.80 \\
0.005    & -305.6457(65)    & 15.29(20)    & 1077(14)    & 0.555(43)    & 0.74 \\
0.006    & -305.6412(79)    & 15.00(17)    & 1018(11)    & 0.552(48)    & 0.69 \\
0.007    & -305.6391(67)    & 14.52(26)    & 1051(53)    & 0.540(52)    & 0.63 \\
0.008    & -305.6530(65)    & 14.72(19)    &  980(10)    & 0.523(56)    & 0.58 \\
0.009    & -305.6555(82)    & 15.28(28)    & 1272(163)   & 0.502(59)    & 0.54 \\
  \hline
{\em Langevin}&            &            &            &            &       \\
0.05    & -305.6417(101)   & 23.13(41)    & 1932(41)    & 0.126(02)    & 0.99 \\
0.1     & -305.6416(68)    & 13.97(22)    & 1189(23)    & 0.240(06)    & 0.97 \\
0.2     & -305.6496(57)    &  9.70(13)    &  812(12)    & 0.408(20)    & 0.89 \\
0.3     & -305.6493(56)    &  9.36(16)    &  817(36)    & 0.487(36)    & 0.78 \\
0.4     & -305.6473(58)    & 12.21(22)    &  834(20)    & 0.485(50)    & 0.61 \\
0.5     & -305.6497(80)    & 17.51(44)    & 1237(52)    & 0.425(58)    & 0.43 \\
  \hline
  \hline
 \end{tabular}
 \caption{The Phenol molecule}
 \label{tab:phenol}
\end{table}

\end{widetext}


\begin{thebibliography}{999}

\bibitem{bressanini1}
Bressanini, D.; Reynolds, P. J. Advances in Chemical Physics 1999,
Wiley, New York, 105, 37.

 \bibitem{C43} S. Chandrasekhar, Stochastic problems in physics and astronomy, {\it Rev. Mod. Phys.} {\bf 15} (1943) 1-89.

\bibitem{metropolis}
Metropolis, N.; Rosenbluth, A. W.; Rosenbluth, M. N.; Teller, A. M.;
Teller, E. J. Chem. Phys. 1953, 21, 1087.


\bibitem{CLS} Canc\`es, E.; Legoll, F.; Stoltz, G. IMA Preprint 2005, \#2040, 
http://www.ima.umn.edu/preprints/apr2005/2040.pdf

\bibitem{umrigar}
Umrigar, C. J. Phys. Rev. Lett. 1993, 71, 408.


\bibitem{MT} Meyn S. P.; Tweedie R. L. Markov Chains and Stochastic
  Stability, 1993, Springer.

\bibitem{caf} Caffarel, M.; Claverie, P. J. Chem. Phys. 1988, 88, 1100.

\bibitem{UNR} Umrigar, C. J.; Nightingale, M. P.; Runge,
  K. J. J. Chem. Phys. 1993, 99, 2865.

\bibitem{sun}
Sun, Z.; Soto, M. M.; Lester Jr., W. A J. Chem. Phys. 1994, 100, 1278.

\bibitem{bressanini2}
Bressanini, D.; Reynolds, P. J. J. Chem. Phys. 1999, 111, 6180.

\bibitem{ceperley}
Ceperley, D.; Chester, G. V.; Kalos, M. H. Phys. Rev. B 1977, 16, 3081.

\bibitem{stedman}
Stedman, N. L.; Foulkes, W. M. C.; Nekovee, M. J. Chem. Phys 1998, 109, 2630.

\bibitem{ricci}
Ricci, A.; Ciccotti, G. Mol. Phys. 2003, 101, 1927.

\bibitem{CJL} Canc\`es, E.; Jourdain, B.; Leli\`evre, T. Cermics,
  Math. Models Meth. Appl. Sciences, in press

\bibitem{bbk}
Br\"unger, B.; Brooks, C. L.; Karplus, M. Chem. Phys. Lett. 1984, 105, 495.

\bibitem{allen}
Allen, M. P.; Tildesley D. J. Computer simulation of liquids, Oxford university
press; and references in section 9.3

\bibitem{stoltz} Izaguirre J.~A.; Catarello D.~P.; Wozniak J.~M., Skeel R.~D.
  J. Chem. Phys. 2001, 114(5), 2090.

\bibitem{michel}
QMC=Chem is a Quantum Monte Carlo program written by M. Caffarel, IRSAMC,
Universit\'e Paul Sabatier -- CNRS, Toulouse, France.

\bibitem{gamess}
Schmidt, M.~W.; Baldridge, K.~K.; Boatz, J.~A.; Elbert, S.~T.; Gordon, M.~S.;
Jensen, J.~H.; Koseki, S.; Matsunaga, N.; Nguyen, K.~A.; Su, S.~J.; Windus, T.~L.;
Dupuis, M.; Montgomery. J.~A. 
J. Comp. Chem. 1993, 14, 1347.

\end{thebibliography}
\end{document}